\documentclass[twocolumn,english]{revtex4-1}
\usepackage[LGR,T1]{fontenc}
\usepackage[latin9]{inputenc}
\usepackage{longtable}
\usepackage{textcomp}
\usepackage{amsmath}
\usepackage{stackrel}
\usepackage{graphicx}
\usepackage{subscript}

\makeatletter

\DeclareRobustCommand{\greektext}{%
  \fontencoding{LGR}\selectfont\def\encodingdefault{LGR}}
\DeclareRobustCommand{\textgreek}[1]{\leavevmode{\greektext #1}}
\ProvideTextCommand{\~}{LGR}[1]{\char126#1}

\providecommand{\tabularnewline}{\\}

\makeatother

\usepackage{babel}
\begin{document}

\title{Experimental development and evaluation of a volume coil with slotted
end-rings coil for rat MRI at 7 T}

\author{S. Solis-Najera\textsuperscript{1}, R. Martin\textsuperscript{1},
F. Vazquez\textsuperscript{1}, O. Marrufo\textsuperscript{2}, A.
O. Rodriguez\textsuperscript{3,}}

\address{\textsuperscript{1}Departamento de Fisica, Facultad de Ciencias, UNAM,
Department of Neuroimage, \textsuperscript{3}Department of Electrical
Engineering, UAM Iztapalapa, DF, 09340, Mexico.}
\begin{abstract}
A volume coil with squared slots-end ring was developed to attain
improved sensitivity for imaging of rat's brain at 7 T. The principles
of the high cavity resonator for the low-pass case and the law of
Biot-Savart were used to derive a theoretical expression of the coil
sensitivity. The slotted-end ring resonator showed a theoretical 2.22-fold
improvement over the standard birdcage coil with similar dimensions.
Numerical studies were carried out for the electromagnetic fields
and specific absorption rates for our coil and a birdcage coil loaded
with a saline-filled cylindrical phantom and a digital brain of a
rat. An improvement of the signal-to-noise ratio (SNR) can be observed
for the slotted volume coil over the birdcage regardless of the load
used in the electromagnetic simulations. The specific absorption rate
simulations show an important decrement for the digital brain and
quite similar values with the saline solution phantom. Phantom and
rat's brain images were acquired at 7 T to prove the viability of
the coil design. The experimental noise figure of our coil design
was four times less than the standard birdcage with similar dimensions,
witch showed a 30\% increase in experimental SNR. There is remarkable
agreement among the theoretical, numerical and experimental sensitivity
values, which all demonstrate that the coil performance for MR imaging
of small rodents can be improved using slotted end-rings.
\end{abstract}
\maketitle

\section{Introduction}

In vivo imaging of mice and rats is soundly established as a component
of preclinical and translational biomedical research {[}1-5{]}. The
biomedical research community has recognized the unique power of magnetic
resonance imaging (MRI) for in vivo measures in small animals {[}6{]}.
The progress of research-dedicated MRI systems equipped with strong
magnets (> 7 T) for small animal imaging has given a boost to RF technology.
The RF receive coil is vital in determining high signal-to-noise ratio
(SNR), and image quality increases with SNR, hence coil selection
is critical for rodent MRI experiments. 

Volume coils for small animal investigations with MRI are a particularly
popular choice for a number of reasons {[}7-8{]}. In particular, birdcage
coils have been a popular design for a number of years. This type
of RF coil offers a convenient geometry because it can generate an
excellent field uniformity, sensitivity, and natural ability to operate
in quadrature. Additionally, they may be placed coaxially with the
bore of the magnet for easy loading and unloading of rodents. The
birdcage coil is still an important subject of study as shown by recent
results {[}8-15{]}.  

The design of dedicated RF coils is key to achieving the best preclinical
and experimental results for MRI. The end rings of the birdcage coil
are an important design aspect as they can modify the intensity and
homogeneity {[}7,10,13-15{]}. The principles of the cavity resonator
proposed by Mansfield et. al. {[}16{]} offer an approach to improve
the intensity and homogeneity of volume coils, with end rings composed
of uniformly distributed slots forming a symmetrical distribution. 

In this paper, we developed a coil design based on the RF coil reported
in {[}17{]} and used for whole-body imaging of rats at 7 T. The coil
designed proposed here is composed of squared slots end rings and
reduced number of rungs to decrease\emph{ }specific absorption rate
(SAR) {[}18{]}. We derived an expression of the sensitivity based
on the low-pass cavity resonator and the law of Biot-Savart to investigate
the coil performance, and experimentally validate this theoretical
frame. This coil design is intended for MR imaging of a rodent's brain.
As we understand, this is the first attempt to experimentally corroborate
the volume coil performance using a specific end ring layout.

\section{Method}

\subsection{End-rings comparison}

There is an important difference between the end-rings in a standard
birdcage coil and the modified design of the cavity resonator presented
here. Fig. 1 shows a schematic and photograph of the coil proposed
in this research.

\noindent\begin{minipage}{1\columnwidth}%
\begin{center}
\includegraphics[scale=0.25]{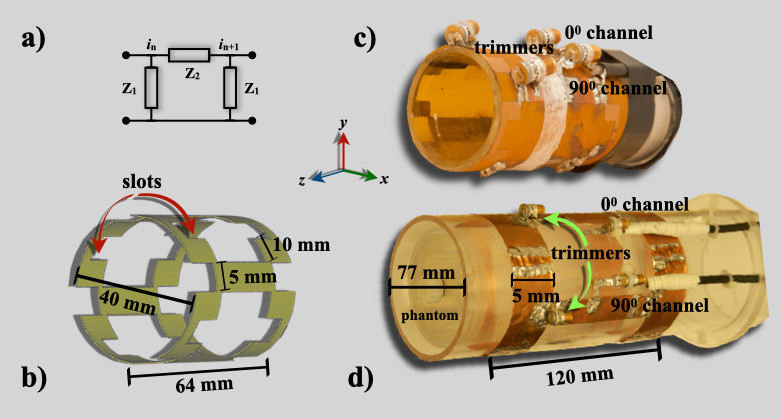}
\par\end{center}
{\small{}Figure 1. a) a $\pi$-section of a lumped parameter transmission
line, b) schematic view of the slotted-end ring volume coil, photographs
of constructed squared-slot end ring coil showing dimensions and passive
electronic components (c), and birdcage coil used for comparison purposes
(d).}{\small\par}%
\end{minipage}

To investigate the effect of the squared-slot end rings, we used the
formalism for the physical principles of the cavity resonator developed
by Mansfield et. al. for the low-pass topology {[}16{]}. From this:

\[
{\displaystyle i^{2}=i_{0}^{2}\frac{R}{r\stackrel[n=1]{N}{\sum}\cos(n\varTheta)}\qquad(1)}
\]

where \emph{R} is the rung resistance and \emph{r} is the slot resistance,
and $\stackrel[n01]{N}{\sum}\cos(n\varTheta)=\frac{N}{2}$, and

\[
R=\frac{2}{Nr}Z_{0}^{2}\qquad(2)
\]

where $Z_{0}$ is the impedance of the slot, and \emph{N} is the number
of rungs. To compute \emph{r}, we use \emph{Q}, the quality factor
of the low-pass cavity resonator:

\[
Q=\frac{2\pi}{Nr}Z_{0}\qquad(3)
\]

Substituting for the resistances \emph{R}, \emph{r} using eqs. (2)
and (3) in eq. (1), we obtain,

\[
{\displaystyle i_{\mathrm{slot}}=i_{0}\frac{Q}{\pi}\qquad(4)}
\]

the ratio $\frac{i\mathrm{_{cav}}}{i_{0}}$ in Eq. (4) does not depend
on the number of rungs, but just the quality factor.

For comparison purposes, we proceeded similarly as above, so the birdcage
end rings can be studied using the intensity of the rung currents
{[}19-20{]}:

\[
{\displaystyle i_{\mathrm{bc}}=i_{0}\frac{1}{2\sin\left(\frac{\pi}{N}\right)}\qquad(5)}
\]

Then, combining eq. (4) and eq. (5):

\[
{\displaystyle i_{\mathrm{slot}}=\frac{2Q}{\pi}\sin\left(\frac{\pi}{N}\right)i_{\mathrm{bc}}\qquad(6)}
\]

In particular, for the 4-rung coil layout,

\[
{\displaystyle i_{\mathrm{slot}}=2.22\,i_{\mathrm{bc}}\qquad(7)}
\]

Once we have computed the currents for the birdcage coil and the cavity
resonator, we can now compare the transverse magnetic field $\mathrm{B}_{1}$.
The $\mathrm{B}_{1}$ at the coil\textquoteright s isocenter relative
to the end-rings current for our coil design with 4 rung is {[}19,20{]}: 

\[
{\displaystyle \mathrm{B}_{\mathrm{\mathit{i}slot}}=2.22\frac{i_{\mathrm{bc}}\left(l^{2}+2d^{2}\right)}{d\left(l^{2}+d^{2}\right)^{3/2}}\qquad(8)}
\]

where \emph{l} and \emph{d} represent the length and the diameter
of the volume coil, respectively, and Eq. (8) was computed with no
shielding. A more detailed derivation of the sensitivity expression
for a birdcage coil can be found in ref. {[}21,22{]}.

\subsection{Electromagnetic field simulations}

The electromagnetic field simulations of RF coils can serve to guide
the development of specific designs for specific applications and
demonstrate how this coil design interacts with the sample to be imaged
{[}23{]}. The commercial software CST Microwave Studio (CST MICROWAVE
STUDIO, CST GmbH, Darmstadt, Germany) was used to calculate the electromagnetic
fields. We have experimentally validated this commercial code with
a birdcage coil for whole-body MRI of rats at 7 Tesla {[}18{]}. 

To numerically calculate the electromagnetic fields of the slotted-end
rings coil, perfect electric conductors (PEC) were assumed, and together
with a four-leg configuration and a saline-solution cylindrical phantom
were used. The phantom properties were $\sigma=5.55\text{\texttimes}10^{\text{\textminus}6}$
S/m, $\varepsilon=78.4$,$\rho=$ 998 kg/m\textsuperscript{3}, and
\textgreek{m} = 0.999991. A 1 V sinusoidal feed was applied, and the
source and conductor impedances were set to 50 \textgreek{W} (pure
resistive). To calculate more realistic results, the rat's brain phantom
($\sigma=0.527133\text{\texttimes}10^{\text{\textminus}6}$ S/m, $\varepsilon=70$,
$\rho=$ 1030 kg/m\textsuperscript{3}, and \textgreek{m} = 1) reported
in {[}18, 24{]} was also used. This rat's brain model is considered
a voxel-based models constructed using digital volume arrays and boundary
representation (BREP) models. These type of models offer an easy implementation
and fast calculation within most commercial simulation codes. {[}25{]}. 

A safety evaluation of RF coils is especially important to protect
the sample from heat and temperature increase {[}26{]}. This is mainly
done by using RF simulations in a heterogeneous body to compute realistic
spatial distributions of SAR. All SAR predictions were computed assuming
1 g averaging, the IEEE STD C-95.3.1-2010 method, 1 W input power
and were performed with open boundary conditions defined in all directions.
Fig. 2.a) and c) show the schematic used in the corresponding numerical
assessments. 

\noindent\begin{minipage}{1\columnwidth}%
\begin{center}
\includegraphics[scale=0.35]{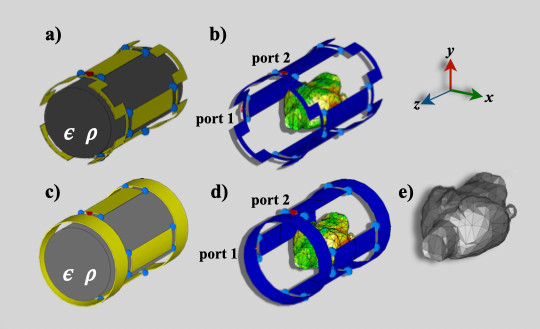}
\par\end{center}
{\small{}Figure 2. The simulation setups for both coils and the cylinder
phantom are shown in a) and c) while the rat's digital phantom is
shown in b) and d). The digital phantom used in the elecromagnetic
and SAR simulations is shown in e).}{\small\par}%
\end{minipage}

\vspace{0.5cm}

The calculations were terminated after simulations of 4.23 pulse width
corresponding to a system energy decay of around - 30 dB for all cases,
and the duration of the excitation was around 7.11 ms. Similarly,
simulations were run for a birdcage with similar dimensions and the
same configurations. The coils were excited in quadrature mode. The
simulation setups, phantoms, and volume coils are shown in Fig. 2.b)
and d).

\subsection{Coil prototype}

To optimize SNR performance, the coil dimensions should match the
size of a mouse while matching the homogeneous RF region. The size
of the volume coil was chosen to accommodate rats while also taking
into account that a 12 cm bore was available. For this study, our
coil design was 64 mm in length, 40 mm in diameter, contained four
rungs, and both end rings were composed of four equally spaced rectangular
slots. Fig. 1.b) and c) show a schematic and a photograph of the coil. 

We designed our coil with four rungs to attenuate the SAR as experimentally
shown by Martin et. al. {[}18{]}. The coil dimensions gives: $\mathrm{diameter/length}=0.625$
to avoid field homogeneity problems and to drastically affect the
SNR. This result is in good concordance with ref. {[}12, 22{]}. Our
coil design is a low-pass resonator because $\frac{\lambda}{20}=50$
cm, where $\lambda$ is the wavelength at 300 MHz. The RF coil prototype
was built on a semi flexible printed circuit board (Pyralux\textsuperscript{®}circuit
flexible material: thickness = 100 $\mu$m, $\epsilon=2.5$, $\tan(\delta)=0.002$.
Dupont\texttrademark , Inc. Wilmington, DE, USA) according to the
specific coil configuration. Wapler et. al. have conducted considerable
investigations on a number of materials suitable to build MRI coils
such as Pyralux\textsuperscript{®}material {[}27{]}. This material
has also been used to print coil arrays for clinical MRI {[}28{]}.
This printed circuit board was mounted on an acrylic cylinder to form
a volume coil. 

The prototype was tuned to 299.47 MHz (the proton frequency at 7 T)
using nonmagnetic chip capacitors and trimmers. One 50 \textgreek{W}-coax
cable was attached to each channel ($0^{0}$ and $90^{0}$ channels)
for quadrature drive, tuning and matching. Rough tuning was achieved
with eleven and six fixed-value chip capacitors (American Technical
Ceramics, series ATC 100 B nonmagnetic) of 3.7 pF and 3.9 pF, respectively.
50-\textgreek{W} matching and fine tuning was achieved using four
nonmagnetic trimmers (Voltronics, Corp: 1-33 pF, NMAJ30 0736) two
for each channel. The resonant frequency for each channel was measured
using a network analyzer (Model 4396A, Hewlett Packard, Agilent Technologies,
CA) as the loss return ($\mathrm{S}_{11}$). 

After fine tuning and matching, both channels were decoupled at the
desired frequencies by altering the balancing capacitor values. The
quality factor (\emph{Q}) of each channel in the coil was also experimentally
determined by measuring the resonant frequency divided by the 3 dB
bandwidth, \textgreek{Dw}, with a quarter-wavelength coaxial cable
at the input of the coil. The loaded \emph{Q} value was measured while
the coil was loaded with a saline-filled spherical phantom (3 cm diameter). 

\subsection{Imaging experiments}

To test the validity of this coil, cylindrical phantom images were
acquired using a standard spin echo sequence. The acquisition parameters
were: TE/TR = 25 ms/900 ms, FOV = 40 mm x 40 mm, matrix size = 256
x 256, slice thickness= 2 mm, NEX = 1. Additionally, images of mouse's
head were acquired using gradient echo sequence and the following
acquisition parameters: TE/TR = 6 ms/400 ms, flip angle = $90^{0}$,
FOV = 35 mm x 35 mm, matrix size = 256 x 256, slice thickness = 1
mm, NEX = 1. 

All MRI experiments were performed on a 7T/21cm Varian imager equipped
with DirectDriveTM technology (Varian, Inc, Palo Alto, CA) and, a
SGRAD 205/120/HD gradient system capable of producing pulse gradients
of 400 mT/m in each of the three orthogonal axes and interfaced to
a VnmrJ 2.1B console. The animal procedures were approved by the Ethical
Committee of UAM Iztapalapa. 

\subsection{Noise factor}

The RF penetration decreases when the coil is filled with a saline-solution
phantom {[}7,20,22{]}. The noise factor (\emph{NF}), is a simple way
to understand the implication of this reduction. If the sample noise
dominates then \emph{NF} can be defined as {[}29{]}:
\[
{\displaystyle NF=20\log_{10}\left[\frac{\mu(\mathrm{B}_{1})-\sigma(\mathrm{B}_{1})}{\sqrt{\frac{1}{n}\sum_{s}\mathrm{B_{1}}(x,y)}}\right]\quad\quad(9)}
\]
where $\mu(\mathrm{B}_{1})$ is the mean and $\mathrm{o}(\mathrm{B}_{1})$
standard deviation, \emph{n} is the number of image voxels, and \emph{s}
is the image space. 

\section{Results and Discussion}

Our coil design was validated by full wave electromagnetic simulations,
theoretical and experimental sensitivity and phantom images. To examine
our coil design in more detail, electromagnetic field simulations
included the electric magnetic field intensity maps of birdcage and
slotted-end ring coils for the following loading cases: a) air-filled,
b) saline solution-filled phantom and c) rat's digital brain phantom.
Fig. 3 shows the bi-dimensional maps of the electric (E) and magnetic
field (B\textsubscript{1}) for the three cases above, and comparison
plots the electromagnetic fields and SNR. 

\noindent\begin{minipage}{1\columnwidth}%
\begin{center}
\includegraphics[scale=0.19]{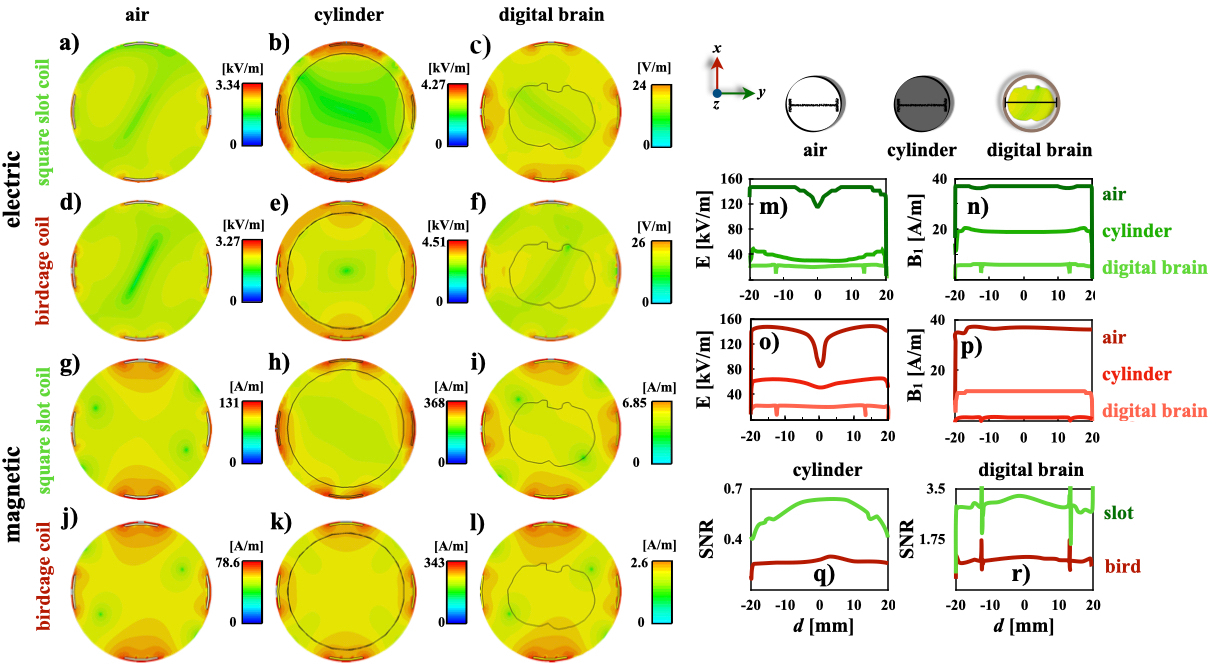}
\par\end{center}
{\small{}Figure 3. Series of bi-dimensional maps of the electric (a-f)
and magnetic (g-l) fields for the birdcage and slotted end-ring coil.
Comparison plots for E (o), B}\textsubscript{{\small{}1}}{\small{}(p)
and SNR for the saline solution (q) and the rat's brain (r) phantoms.}{\small\par}%
\end{minipage}

\vspace{0.5cm}

The bi-dimensional maps of the magnetic fields of both volume coils
have very strong similarities, as shown in Fig. 3. g)-l). The simulated
electromagnetic field results of both coil designs for the saline
solution phantom corroborate to: a) those results reported by Webb
{[}30{]}, b) numerical evaluations of a band-pass-birdcage coil (ratio
= 1 and 8 rungs) computed with a finite element modeling at 127.74
MHz {[}31{]}, c) a quadrature low-pass birdcage coil (ratio = 0.7
and 32 rungs) obtained at 7 T for small animal MRI {[}32{]} (coil
dimensions are essentially the same as ours in this study), d) numerical
results obtained using the finite-difference time-domain method for
a low-pass birdcage coil (ratio = 0.7 and 8 rungs) driven in quadrature
mode at 7 T for rodents {[}33{]}, e) a quadrature birdcage coil (ratio
= 0.88 and 16 rungs) for MRI of rabbits at 7 T {[}34{]}, and f) various
birdcage coils (ratio = 0.62 and 8 rungs) with different rung cross
sections at 9.4 T {[}35{]}. 

Additionally, the B\textsubscript{1} pattern of the saline solution
simulations for both coil designs (Fig. 3.n) shows good concordance
with simulations obtained at 200 MHz and 400 MHz using a simple bi-dimensional
full wave model developed by Spence and Wright {[}36{]}, and similar
values of B\textsubscript{1} and field pattern were reported by Doty
et. al. for a 2.5 cm Litzcage at 300 MHz {[}37{]}. The bi-dimensional
representations of the electric field for both coil designs in Fig.
3.a)-c) and g)-i) are able to produce the expected behaviour as reported
in {[}20,31{]}. The patterns produced by both coils are quite comparable
regardless of whether the coil is filled with a phantom or not. 

However, the air-filled coil case shows the best agreement. The pattern
of the electric field reported by Kangarulu et. al. {[}38{]} for a
transverse electromagnetic (TEM) coil with resonant frequency of 340
MHz confirms the patterns of the electric field of both coil designs
in Fig. 3. m) and n). Following the simulated data of the electromagnetic
fields and the methodology proposed in {[}39{]}, we calculated plots
of the electric field as a function of the $\mathrm{B}_{1}$ field. 

Fig. 4 illustrates the plots for the saline solution and digital brain
phantoms. As expected, from this we can observe that there is a linear
relation between the E and $\mathrm{B}_{1}$, and the slope magnitudes
are very similar to the experimental results obtained using an electro-optic
probe at 128 MHz and 200 MHz. Another important fact is that regardless
of the type of phantom used the slope value is practically the same.
These experimental results validate the simulations of the electric
field produced by the slotted end-ring resonator.

\vspace{0.01cm}

\noindent\begin{minipage}{1\columnwidth}%
\begin{center}
\includegraphics[scale=0.4]{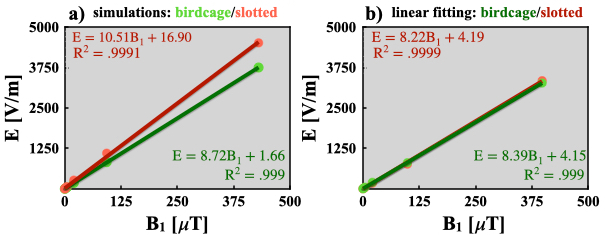}
\par\end{center}
{\small{}Figure 4. The linear relation between the E and B}\textsubscript{{\small{}1}}{\small{}
magnitudes of the birdcage coil and the slotted end-ring resonator
for: a) saline solution phantom and b) digital brain phantom. }{\small\par}%
\end{minipage}
\vspace{0.01cm}

Profiles of Fig. 3.m) and o) show that the electric field of the
birdcage coil has a greater intensity when compared to our coil design.
It is important to mention that a great deal of electrical energy
was absorbed by both phantoms as shown in Fig. 3.m): when the coil
is empty, the greatest energy levels were obtained, as indicated by
the green profile, and for the rat's brain phantom profile, an even
greater amount of energy is absorbed. 
The simulated SNR was computed using the bi-dimensional maps of the
$\mathrm{B}_{1}$ and E. Comparison plots were computed and shown
in Fig. 3. q) and r). These SNR profiles depict a clear numerical
improvement of the slotted end-ring coil over the standard birdcage
coil for both cases. Numerical modeling of the interaction between
the electromagnetic fields and the animal model provides a useful
way to assess the rate of energy deposition. The $\mathrm{SAR_{1g}}$
(1 g averaging SAR) numerical assessments for the saline solution
and digital brain phantoms are shown in Fig. 5.

Additionally, to compare the $\mathrm{SAR_{1g}}$ results, comparison
histograms and profiles were obtained for the both phantoms and the
standard orientations. The $\mathrm{SAR_{1g}}$ bi-dimensional maps
of Fig. 5.a)-f) show a very good concordance with bi-dimensional maps
of SAR obtained via $\mathrm{B}_{1}$ mapping of a quadrature birdcage
coil (diameter 60 cm) loaded with solution-filled cylinders at 64
MHz T {[}40{]}. These $\mathrm{SAR_{1g}}$ predictions in the axial
(Fig. 5.a) and d)) and the coronal (Fig. 5.b) and e)) orientations
agree very well with results obtained via the tomographic method and
a birdcage coil (ratio = 0.91 and 16 rungs) tuned a 128 MHz {[}41{]}
as well as phantom imaging studies conducted by Cline et. al. {[}42{]}.

\noindent\begin{minipage}{1\columnwidth}%
\begin{center}
\includegraphics[scale=0.36]{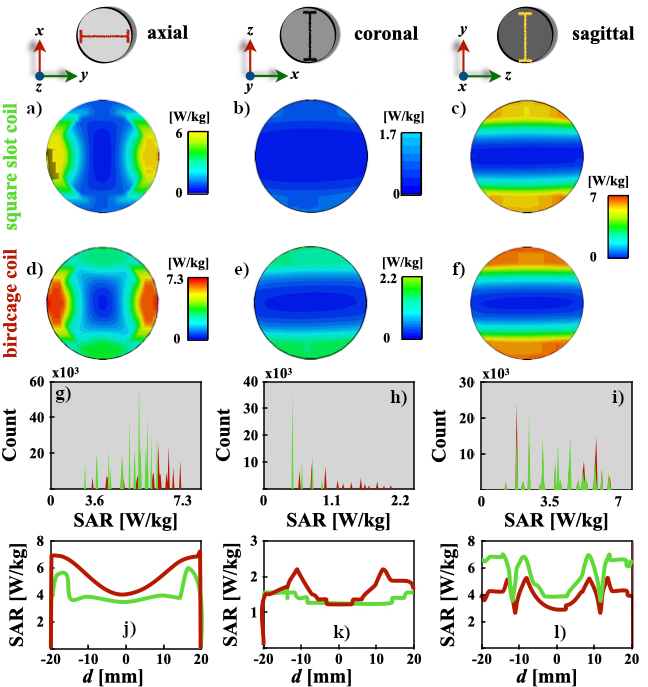}
\par\end{center}
{\small{}Figure 5. Series of bi-dimensional maps of $\mathrm{SAR_{1g}}$
for the two volume coils (a)-(f). Simulation data were used to compute
for comparison histograms (g)-(i) and $\mathrm{SAR_{1g}}$ plots (j)-(l)
as indicated in the top row illustrations.}{\small\par}%
\end{minipage} 
\vspace{0.5cm}

There is a reasonable agreement between the theoretical results published
by Hoult for a quadrature-driven volume coil at 200 MHz {[}43{]},
and the comparison plot for axial orientation of Fig. 5.j). Histograms
in Fig. 5.g)-i) show that the birdcage coil has a distribution with
higher values of $\mathrm{SAR_{1g}}$ compared to our coil design
for all three directions. The coronal cut histogram of Fig. 5.g) has
a wider distribution of significantly higher values for the birdcage
coil, and the slotted end-ring coil has a much lower rate of absorbed
energy for the same interval. Fig. 5.h) shows that the absorbed energy
ratio by the slotted volume coil is less than 1.1 W/kg, and the birdcage
coil shows higher values for a wider interval approximately between
0.55 and 2 W/kg. The Fig. 4.i) distribution of energy absorbed rate
by the phantom looks roughly the same along the same interval, which
is confirmed by the comparison plot in Fig. 5.l). Comparison plots
of $\mathrm{SAR_{1g}}$ in Fig. 5.j)-l) show practically the same
pattern and intensity for both coil designs. edit: However, the birdcage
coil shows a slight increase over the slotted end-ring coil in the
axial and sagittal orientations; see Fig. 5.j) and l). This is more
easily appreciated at both ends, while towards the centre, the $\mathrm{SAR_{1g}}$
intensity tends to be roughly the same. Similarly, numerical assessments
of $\mathrm{SAR_{1g}}$ for the rat's digital brain phantom were computed
to obtain more realistic results. 

Fig. 6 shows results of $\mathrm{SAR_{1g}}$ for three different orientations.
These bi-dimensional maps correspond very well with simulated (100
mg averaging) $\mathrm{SAR_{100mg}}$ predictions reported by Martin
et. al. {[}18{]}. The $\mathrm{SAR_{1g}}$ intensity values agree
with results reported by Wang et. al. {[}44{]}, which assumes electromagnetic
plane waves and uses the (Finite-Difference Time-Domain) FDTD method.
\vspace{0.5cm}

\noindent\begin{minipage}{1\columnwidth}%
\begin{center}
\includegraphics[scale=0.31]{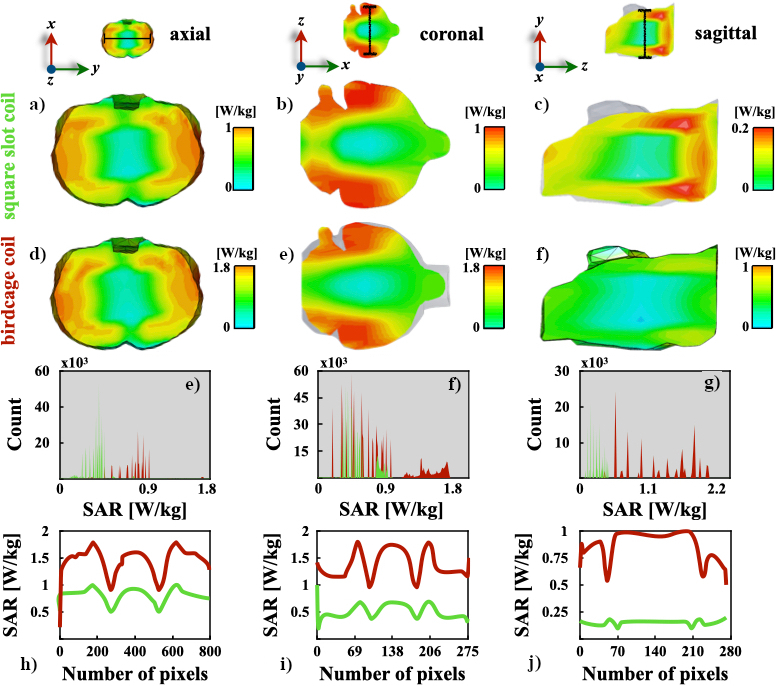}
\par\end{center}
{\small{}Figure 6. Series of bi-dimensional maps of $\mathrm{SAR_{1g}}$
for the two volume coils (a)-(f). Simulation data were used to compute
for comparison histograms (g)-(i) and $\mathrm{SAR_{1g}}$ plots (k)-(l).}{\small\par}%
\end{minipage}
\vspace{0.5cm}

The $\mathrm{SAR_{1g}}$ and absorbed power results obtained from
simulations in Fig. 6 are summarised in Table 1.

\onecolumngrid
\noindent \begin{center}
{}%
\begin{longtable}{|c|c|c|c|c|}
\hline 
 & \multicolumn{2}{c|}{\textbf{Birdcage coil}} & \multicolumn{2}{c|}{\textbf{Slotted-end ring coil}}\tabularnewline
\hline 
\textbf{Phantom type/}{${\scriptstyle \mathrm{SAR_{1g}}}$} & {Saline solution} & {Digital brain} & {Saline solution} & {Digital brain}\tabularnewline
\hline 
{Absorbed power {[}mW{]}} & {179.30} & {9.45} & {147.45} & {3.66}\tabularnewline
\hline 
{total ${\scriptstyle {\scriptscriptstyle \mathrm{SAR_{1g}}}}$
{[}W/kg{]}} & {2.99} & {1.20} & {2.46} & {0.46}\tabularnewline
\hline 
{max${\scriptstyle \mathrm{{\scriptscriptstyle SAR_{1g}}}}$
{[}W/kg{]}} & {7.95} & {1.79} & {8.22} & {0.51}\tabularnewline
\hline 
{max${\scriptscriptstyle \mathrm{SAR_{1g}}}$/total ${\scriptscriptstyle \mathrm{SAR_{1g}}}$} & {2.65} & {1.49} & {3.34} & {1.10}\tabularnewline
\hline 
{max$\mathrm{{\scriptscriptstyle {\scriptstyle SAR_{1g}}}}$
location (\emph{x,y,z}) {[}mm{]}} & {(0.42, -16.50, 17)} & {(13.15, -1.26, 26) } & {(0.46, -16.51, 47.33)} & {(-3.25, -8.37, 15.61) }\tabularnewline
\hline 
\end{longtable}{\tiny\par}
\par\end{center}

\begin{center}
{\small{}Table 1. Comparison of $\mathrm{SAR_{1g}}$ results and the
locations of max$\mathrm{SAR_{1g}}$ }{\small\par}
\par\end{center}
\twocolumngrid

$\mathrm{SAR_{1g}}$ predictions of the rat's brain phantom for the
slotted volume coil are lower than 0.5 W/kg in the axial and sagittal
orientations as shown in histograms of Fig. 6.e) and g). In the coronal
cut of Fig. 6.f), the $\mathrm{SAR_{1g}}$ values are lower than 0.9
W/kg, and the birdcage coil produces a higher ratio for the entire
interval. Comparison plots of Fig. 6.h)-j) clearly show lower $\mathrm{SAR_{1g}}$
values of the slotted volume coil than the ones obtained with the
birdcage coil for all three cuts. However, the highest difference
is in the saggital direction as illustrated in Fig. 6.j). The simulated
absorbed power values of both coil are in reasonable concordance with
the theoretical values for coil radii: $r_{\mathrm{slotted}}=2.28$
cm and $r_{\mathrm{birdcage}}=2.58$ cm, obtained with the analytical
model of birdcage resonators at similar resonant frequencies and reported
by Foo et. al. {[}45{]}. The $\mathrm{SAR_{1g}}$ bi-dimensional mappings
of Fig. 6.a)-f) and their corresponding histograms (Fig. 6.e)-g) confirm
the simulated results obtained with a digital anatomical model of
the Sprague-Dawley rat (voxel dimension $1.95\,\mathrm{x}\,1.95\,\mathbf{x}\,2.15\,\mathrm{mm}^{3}$),
based on MRI data and the FDTD numerical approach {[}46{]}. These
calculations show that the slotted volume coil has better agreement
with this analytical model despite our brain model having a much lower
resolution ($0.18\,\times\,0.18\,\times\,0.5\,\mathrm{mm}^{3}$) {[}18{]}. 

Greater power absorption can be observed for the saline solution phantom
when compared to the digital brain phantom for both coils: an approximately
19-fold (birdcage coil) and 40-fold (slotted volume coil) increase.
So, the slotted volume coil is able to absorb four times more energy
than the birdcage coil. As expected, the use of digital phantoms of
specific organs provide more realistic results, despite the fact that
our BREP brain phantom is not able to reproduce accurately complex
anatomical details. The numerical assessments of the saline solution
phantom $\mathrm{SAR_{1g}}$ are practically the same for both coil
designs. This is expected because both coil designs have similar topologies.
However, the slotted volume coil produces a much lower $\mathrm{SAR_{1g}}$
value when using the digital brain phantom. Something very similar
happens for the max$\mathrm{SAR_{1g}}$ calculations which are also
in good agreement with values obtained using a probabilistic approach
{[}47{]}. From the max$\mathrm{SAR_{1g}}$ results, we can observe
that both coils have roughly the same values for the saline solution
phantom, and when the digital brain model was used as a load, a 3.5-fold
increase was produced. These max$\mathrm{SAR_{1g}}$ computations
corroborate very well with those computed by Trakic et. al. {[}48{]}
using a Sprague Dawley rat model and a birdcage resonator (ratio =
1) operating at 500 MHz. The location of max$\mathrm{SAR_{1g}}$ values
are in the same quadrant with exception of the case for the slotted
volume coil and the digital brain phantom. 

The noise figures of both coil prototypes were computed according
to eq. (9), giving $NF\mathrm{_{slot}}\approx1$, and $NF\mathrm{_{bc}}\approx4$.
These two noise figure values have an adequate concordance with those
theoretical results for half birdcage and U-shaped split birdcage
resonators reported by Gasson et. al. {[}29{]}. This important reduction
of \emph{NF}: $NF\mathrm{_{bird}}\approx4NF\mathrm{_{slot}}$, is
due to the size of the birdcage resonator and eq. (8). 

To characterise our coil design, the S-parameters and the Smith chart
were experimentally measured. The S-parameter plots and Smith charts
of both channels are in Fig. 7. Additionally, S$_{11}$, S$_{12}$
and S$_{21}$ parameters were numerically and experimentally computed
for the slotted volume coil of Fig. 1.c). This S-parameter comparison
has a major concordance between numerical simulations and experimental
bench testings. Thus, reliable conditions can be obtained to guide
the design of new RF resonators {[}18, 49{]}. 
\vspace{0.5cm}

\noindent\begin{minipage}{1\columnwidth}%
\begin{center}
\includegraphics[scale=0.2]{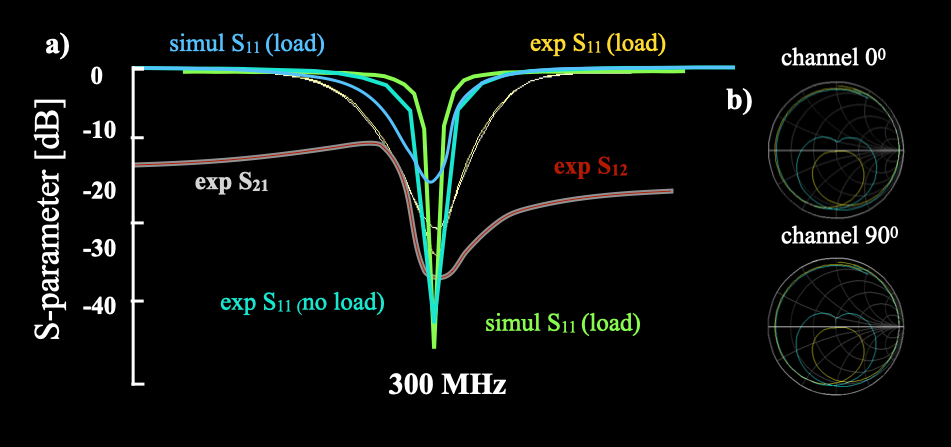}
\par\end{center}
{\small{}Figure 7. a) S-parameters of the slotted volume coil comparison
before and after loading for the simulation and experimental case.
b) Smith charts under the same conditions as in (a): blue line with
load and yellow line without load. }{\small\par}%
\end{minipage}
\vspace{0.5cm}

Both channels showed a good RF penetration and impedance values. Impedance
values are in good agreement with values reported by {[}50{]} and
provides an optimal energy transfer. These parameters show a value
lower than -20 dB, confirming a good 50-\textgreek{W} match and low
decoupling between the two channels to drive the coil in quadrature
mode {[}30{]}. So, these S\textsubscript{11}-parameters profiles
show a good isolation of both channels to assure optimal energy transmission
and reception of the RF signals. These results are in very good agreement
with those reported in {[}22{]}. All the bench testing results are
summarised in Table 2. 
\begin{center}
{\footnotesize{}}%
\begin{tabular}{|c|c|c|c|}
\hline 
{\small{}channel} & \multicolumn{1}{c|}{\emph{\small{}Q}{\small{} factor}} & \multicolumn{1}{c|}{{\small{}RF penetration {[}dB{]}}} & \multicolumn{1}{c|}{{\small{}Impedance {[}$\Omega${]}}}\tabularnewline
\hline 
\multicolumn{4}{|c|}{{\small{}loaded/unloaded}}\tabularnewline
\hline 
{\small{}$0^{0}$} & {\small{}9.6/12.49} & {\small{}-35.84/-44.63} & {\small{}51.42/49.69}\tabularnewline
\hline 
{\small{}$90^{0}$} & {\small{}8.1/11.9} & {\small{}-38.63/-40.95} & {\small{}51.30/49.83}\tabularnewline
\hline 
\end{tabular}{\footnotesize\par}
\par\end{center}

\begin{center}
{\small{}Table 2. Bench testing values for both channels of our volume
coil prototype.}{\small\par}
\par\end{center}

These \emph{Q} values are in very good concordance with those reported
by Marrufo et. al. {[}18{]}. The slotted end-ring coil shows a slightly
better performance than a similar coil previously published with larger
dimensions {[}17{]}. 

Phantom and rat's head images were also acquired: Fig. 8.a) and b)
shows a comparison of axial images of the spherical phantom acquired
with a birdcage coil and our coil, respectively. Profiles of B\textsubscript{1}
magnitude were computed using the simulation, theoretical (Eq. (8))
and experimental results, and a comparison plot was computed, see
Fig. 8.d). Comparison of uniformity and histograms for both constructed
coils were also calculated using the image data of Fig. 8.a) and shown
in Fig. 8.e) and c). All profiles produced an important concordance
to experimentally validate the simulation and theoretical results.
The profile patterns also show an excellent agreement with those reported
for a birdcage coil tuned and matched to 128 MHz {[}51{]}. 

The histogram of the slotted end-ring coil clearly shows a better
performance over the traditional birdcage coil. 
\vspace{0.5cm}

\noindent\begin{minipage}{1\columnwidth}%
\begin{center}
\includegraphics[scale=0.5]{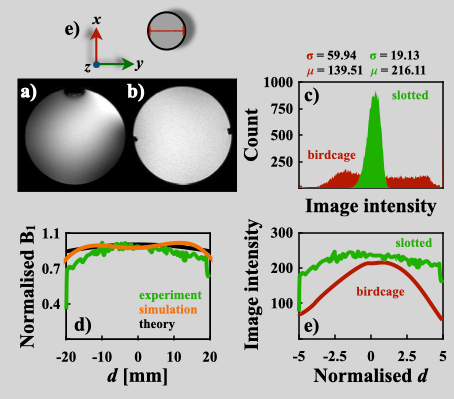}
\par\end{center}
{\small{}Figure 8. Phantom images: (a) birdcage coil (in-plane resolution
= 312.5 $\mu$m x 312.5 $\mu$m x 1 $\mathrm{mm^{3}}$) and, (b) slotted
end-ring resonator (in-plane resolution = 117.2 $\mu$m x 117.2 $\mu$m
x 1 $\mathrm{mm^{3}}$), c) comparison of uniformity plot for simulation,
theoretical and experimental results, and d) comparison of experimental
data for the birdcage coil prototype of Fig. 1.d) and the slotted
end-ring coil. All profiles were taken along the red line in (e).}{\small\par}%
\end{minipage}
\vspace{0.5cm}

The SNRs were also calculated using the image data of Fig. 8.a) and
b). The SNR values for the coils were approximately 23 (slotted end-rings
coil) and 17.6 (birdcage coil). The coil design proposed here is able
to produce a reasonable improvement on performance over a standard
birdcage coil. Consequently, the phantom image acquired with our coil
prototype shows a better quality image and good uniformity compared
to the image obtained with the birdcage coil. Successful ex vivo results
of a rat\textquoteright s brain were obtained with our volume coil
prototype at 300 MHz and shown in Fig. 9. These images of the Wistar
rat head show specific brain structures with a high signal intensity,
excellent image uniformity, and no movement artifacts. The phantom
and rat's head images prove the compatibility of the slotted end-ring
resonator with standard pulse sequences at 300 MHz. 
\vspace{0.5cm}

\noindent\begin{minipage}{1\columnwidth}%
\begin{center}
\includegraphics[scale=0.5]{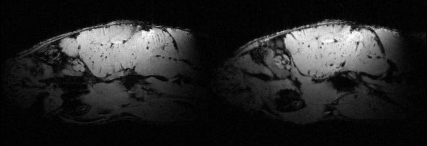}
\par\end{center}
{\small{}Figure 9. In vivo measurements (in-plane resolution = 117.2 $\mu$m
x 117.2 $\mu$m x 1 $\mathrm{mm^{3}}$) using the slotted end-ring
coil and standard spin echo sequences. The anatomical structures of
the rat's brain can be clearly identified.}{\small\par}%
\end{minipage}

\section{Conclusion}

An RF volume coil with a slotted end-ring was presented. The important
parameters of this coil prototype have been calculated analytically
and numerically, and confirmed experimentally on the bench and with
the MRI. The slotted end-ring coil demonstrates better coil performance
than the birdcage coil in terms of sensitivity, homogeneity, SNR and
SAR reduction. We have demonstrated that using full-wave electromagnetic
simulations and experiments in the Varian 7 T MR imager, the slotted
end-ring coil can outperform the birdcage resonator for MRI of small
rodents. This can prove to be advantageous when performing functional
MRI of rats' brain where the coil performance plays an important role
in acquiring optimal MR signals. 

\section{Acknowledgments. }

We thank CONACYT Mexico for research grant 112092. We would like to
thank Mr. Patrick Geeraert for English proofreading.email: arog@xanum.uam.mx.

\end{document}